\documentclass{article}

\begin{document}

\title{MAGNETIC FIELD MEASUREMENT IN BLACK HOLE X-RAY BINARY
CYGNUS X-1\\ {\small (Submitted in {\it Nature})}}

\author{E.A. Karitskaya$^1$, N.G. Bochkarev$^2$,
S. Hubrig$^3$, Yu.N. Gnedin$^4$,\\ M.A. Pogodin$^4$, R.V.
Yudin$^4$, M.I. Agafonov$^5$, O.I. Sharova$^5$ \medskip \\
 {\small (1) Astronomical Institute of RAS, 48 Pyatnitskaya
 str., Moscow, 119017, Russia, karitsk@sai.msu.ru.}\\
 {\small (2) Sternberg Astronomical Institute 13 Universitetskij
 pr., Moscow, 119991, Russia.}\\
 {\small (3) ESO, Chile.}\\
 {\small (4) Central Astronomical Obseravtory at Pulkovo RAS,
 St.-Petersburg, 196140, Russia.}\\
 {\small (5) Radiophysical Research Institute (NIRFI), 25/12a
 B.Pecherskaya str., Nizhny Novgorod, 603950, Russia.}}

\maketitle

\begin{abstract}
X-ray binary Cygnus X-1 is microquasar containing historically
first candidate to black hole. Paradigm of magnetic disc accretion
dominates in theoretical models describing processes taking place
in objects containing black holes such as microquasars and active
galactic nuclei. Nevertheless up to now there were no reliable
measurements of magnetic fields in these systems. The first
prediction of the Cyg X-1 magnetic field was done by
V.F.Shvartsman $^{1}$. He wrote that Cyg X-1 X-ray emission
millisecond flickering evidences the presence of a black hole and
points to the magnetic field role in accretion onto a black hole
$^{2,3}$. From that times there were many of attempts to search
for the Cyg X-1 magnetic field but all of these efforts indicated
upper limits only. Our VLT FORS1 2007 and 2008 observations
revealed a presence of a magnetic field in the system. For the
first time we obtained on the level of 6 standard deviations
($\sigma )$ a magnetic field of the order of 130 G on the surface
of the Cyg X-1 optical component (O-supergiant) and observational
estimation on a 4$\sigma $ level of a magnetic field on a outer
part of accretion structure (about 600 G) in accordance with
theoretical prediction. Scaling this field value to black hole
vicinity we showed that the field is strong enough to explain the
X-ray millisecond flickering and can explain it. Our result
presents the first direct determination of magnetic field in
accreting disk around a black hole.
\end{abstract}

\bigskip

\begin{table}
\caption[]{Magnetic fields from VLT spectropolarimetrical observations
 of X-ray binary Cyg X-1. JD - Julian date of the middles of the object
 observations; Orbital phases $\phi$, according ephemeredes$^{22}$,
 $\phi = 0$ - optical component is ahead; $\sigma$ - standard deviation,
 $|\langle B_z\rangle /\sigma |$ - significance.}
 {\small
\begin{center}
\begin{tabular}{|c|c|c|c|c|c|c|c|c|}
 \hline
 & & \raisebox{-1.50ex}[0cm][0cm]{\textbf{Orbital}} & \multicolumn{3}{|c|}{\textbf{Optical component}} &
 \multicolumn{3}{|c|}{\textbf{Outer part of}} \\
 \raisebox{-0.50ex}[0cm][0cm]{\textbf{Date }} &
 \raisebox{-0.50ex}[0cm][0cm]{\textbf{JD}} &
 \raisebox{-1.50ex}[0cm][0cm]{\textbf{phases}} &
 \multicolumn{3}{|c|}{\textbf{(O-star)}} &
 \multicolumn{3}{|c|}{\textbf{accretion structure}}\\
 \cline{4-9} & & &
 \textbf{$\langle B_z\rangle$, G}& \textbf{$\sigma $, G}& \textbf{$|\langle B_z\rangle /\sigma |$}& \textbf{$\langle B_z\rangle$, G}& \textbf{$\sigma $, G }& \textbf{$|\langle B_z\rangle /\sigma |$} \\
 \hline 18-19 June '07& 2454270.768& 0.650& -6& 28& -0.2& -780& 177& 4.4 \\
 \hline 19-20 June '07& 2454271.778 & 0.830& 37& 22& 1.7& 5& 104& 0.05 \\
 \hline 20-21 June '07& 2454272.760 &0.006& 58& 21& 2.8& 126& 174& 0.72 \\
 \hline 25-26 June '07& 2454277.808 & 0.907& 22& 28& 0.8& -257& 222& -1.15 \\
 \hline 29-30 June '07& 2454281.707 & 0.603& 48& 20& 2.4& 235& 260& 0.90 \\
 \hline  9-10 July '07& 2454291.766 & 0.400& 101& 18& 5.5& 128& 89& 1.4 \\
 \hline 14-15 July '08& 2454662.711 & 0.641& 49& 23& 2.1& -260& 221& -1.2 \\
 \hline 15-16 July '08& 2454663.684& 0.816& 22& 22& 1.0& -95& 122& -0.78 \\
 \hline 16-17 July '08& 2454664.692& 0.995& 80& 23& 3.5& -380& 231& -1.6 \\
 \hline 17-18 July '08& 2454665.692& 0.174& 24& 19& 1.3& 25& 117& 0.21 \\
 \hline 23-24 July '08& 2454671.704& 0.247& -16& 20& -0.8& -93& 82& -1.1 \\
 \hline 24-25 July '08& 2454672.728& 0.430& 27& 19& 1.4& 449& 112& 4.0 \\
 \hline 30-31 July '08& 2454678.676& 0.500& 128& 21& 6.2& 56& 189& 0.29 \\ \hline
\end{tabular}
\label{tab1}
\end{center}
}
\end{table}

Cyg X-1/HDE226868 is a X-ray binary system with the orbital period
P=5.6$^{d}$, whose relativistic component is the first candidate
black hole (BH). The optical component (O9.7 Iab supergiant) is
responsible for about 95{\%} of the system optical luminosity. The
remaining 5{\%} are due to the accretion structure (disc and
surrounding gas) near BH.

Though the investigation of Cyg X-1 are being carried out over 40
years and have resulted in $\sim $1000 publications, geometrical
and physical parameters the system remain unclear. The same is
also true for some phenomena observed in this system including
e.g. long time periodic and aperiodic variations and flares.

Both linear and circular polarizations was detected in Cyg X-1
optical continuum and investigated in 1970s. Linear interstellar
and circumstellar polarization reaches a value of $\sim $5{\%}
$^{4}$. Its strength and position angle change on the scale of
years $^{5}$. Kemp $^{6}$ detected a component of an amplitude
$\sim $0.25{\%} which is variable over the orbital phase $\phi $.
This component shows complicated and variable dependence from
$\phi $. Similar behaviour was found for circular polarization
discovered in 1972 $^{7,8}$. Interstellar circular polarization
does not exceed 0.04{\%}, and variable component with the period
of 2.8/5.6$^{d}$ is about 0.02{\%} $^{9}$. While intrinsic
circular polarization is most probably generated by a magnetic
field, the intrinsic linear polarization in Cyg X-1 is usually
explained by electron scattering on non-symmetrical gas structures
(e.g. $^{10-13})$.

Theoretical estimations of the strength of the magnetic field in
Cyg X-1 were based on optical polarization $^{14, 15}$. Upper
limits $B < 350$ G for the optical component and $B < 500$ G for
the outer part of accretion disc were found. The 6-m telescope
spectropolarimetric observations show $B < 1000$ G for the outer
part of the disc $^{16}$.

Our method of the study of the presence of a magnetic field is
based on measurements of the circular polarization (the Stokes
parameter V) in optical spectra produced by Zeeman effect. The
method of the determination of the mean (averaged over the picture
plane) longitudinal magnetic field $\langle B_{z}\rangle$ is
described in full detail by $^{17}$. The method is statistical: to
increase the sensibility there are used simultaneously all
observed spectral lines. Value $\langle B_{z}\rangle$ is obtained
$^{17-19}$ from the slope of least squares linear regression of

\[
Y = V/I
\]

and

\[
X=-4.67\times 10^{-13}\times g_{eff}\times \langle B_z\rangle
\times \lambda ^2 \times (1/I)\times dI/d\lambda .
\]

Here I is intensity of radiation, $\lambda$ is wavelength in
angstroms, $g_{eff}$ is effective Lande factor.

Our spectropolarimetrical observations were conducted with the
European South Observatory Very Large Telescope (VLT) 8.2 m (Cerro
Paranal, Chile) in service mode with the FORS1 spectrograph in the
range 3680-5129 {\AA}, spectral resolution R=4000, signal-to-noise
ratio S/N =1500 -- 3500 (for spectra of intensity) in 2007 from
June 18 to July 9 and in July, 2008 (see Table 1). The system Cyg
X-1 was at that time in its X-ray ``hard state''. 13
spectropolarimetric spectra with exposure time of $\sim $1 hour
were obtained during 13 nights. For our observations we adopt
effective Lande-factor g$_{eff}$=1.07 according $^{20}$.

The used method has been already applied (and carefully tested) in
previous studies of bright magnetic stars $^{17-19}$ which usually
do not have significant interstellar or intrinsic linear
polarization and have rather strong $\langle
B_{z}\rangle$=500--2000 G. In contrast, Cyg X-1 has $\langle
B_{z}\rangle$ weaker and strong interstellar / circumstellar
linear polarization. For this reason we had to meet some
precautions in the magnetic field measurements of this system and
to adapt the method for such conditions.

Before $\langle B_{z}\rangle$ calculations we clean carefully V/I
spectra from any features which could distort results. We excluded
features alien to photosphere of Cyg X-1 optical component: 1) the
wavelengths of interstellar lines; 2) defects (including weak
residual cosmic ray tracks remained after standard observation
processing); 3) HeII 4686 emission line and 4) emission components
of the lines with strong P Cyg effect. We did not found the
pollution by telluric lines in our spectra.

The continua of the recorded V/I spectra show slight slopes. The
continuum level drop 0.05{\%} - 0.15{\%} within our spectral range
3680-5129 {\AA} varied from night to night. This behaviour cannot
be explained by Cyg X-1 interstellar or/and intrinsic circular
polarization. The probable reason could be a presence of a
cross-talk between linear and circular polarization within the
FORS1 analysing equipment. In agreement with previous studies of
Cyg X-1 linear polarization in optical range $^{5}$ this effect is
observed only in the continuum and does not distort S-shape V
profiles of spectral lines caused by Zeeman effect (hereinafter
Zeeman S-waves).

To avoid any impact of the continuum slope on our $\langle
B_{z}\rangle$ measurements, we subtracted linear trends from V/I
spectra. After removal of slopes the $\langle B_{z}\rangle$
becomes lower by 20-80 G depending on the slope values (all
$\langle B_{z}\rangle$ corrections are negative).

We normalized I-spectra by pseudo-continuum following $^{18, 19}$.
I-continuum is produced by the source energy distribution,
interstellar reddening, broad diffuse interstellar bands (DIBs) as
well as atmospheric extinction and detector sensitivity. Its slope
reaches $\vert d(log(I(\lambda ))/d(log(\lambda ))\vert \sim 20$.
The slope removing gives $\langle B_{z}\rangle$ correction up to
$\sim 20$ G. It is usually less than the statistical errors
$\sigma (\langle B_{z}\rangle)\sim 20-30$ G, see table 1.

After abovementioned reductions the residual deviations of least
squares linear regression become to follow to the Gauss function
satisfactorily up to $\pm $3.6 $\sigma $(V/I), where $\sigma
$(V/I) is standard deviation of V/I. It is mean that the level of
significance corresponds to Gauss statistics now. Small number of
points has larger deviations. Weak cosmic ray tracks can be a
source of such deviations. Therefore we excluded from our analysis
any pixel showing residual deviations exceeding $3.6\times \sigma
(V/I)$.

The results of our measurements are presented in table 1 and
figure 1. To verify our results we used some tests: 1) each
spectrum was divided in two halves at mid-wavelength; we checked
that $\langle B_{z}\rangle$-values determined over each half
separately were in agreement within error-bars. 2) We repeated
$\langle B_{z}\rangle$ calculation using fragments of spectra,
which include strong absorption lines (deeper than 4{\%}) only;
$\sim $1/3 spectral points were used; $\langle B_{z}\rangle$-
values were consistent with the previous measurements using the
whole spectral regions within 1.5 sigma ($\langle B_{z}\rangle)$.
3) Zeeman S-waves for the strongest lines were found. In Fig. 2a
we show an example of a distinct Zeeman feature in the HeI line at
$\lambda $4026{\AA}.

We should note that the element overabundance on the factor from 2
to 10 in the Cyg X-1 optical component stellar atmosphere $^{22-23
}$enforces the spectral lines and Zeeman S-waves. It increases
accuracy of $\langle B_{z}\rangle$ measurements.

As a next step of our study we investigated the spectral line
HeII$\lambda $4686{\AA} separately. Due to the presence of a
strong emission component in the line profile it was omitted from
the earlier analysis. In fact, this line has compound profile
consisting of absorption (originating in the stellar photosphere)
and emission (originating in the accretion structure) components.
Certainly, the accuracy of the measurement of the magnetic field
using just one line is considerably worse compared with
measurement using the whole spectrum. Nevertheless, our analysis
shows that for two spectra accuracies of estimations exceed
4$\sigma $ level: $\langle B_{z}\rangle$=-780+/-177 G in 2007 for
the orbital phase $\varphi $=0.65 and $\langle
B_{z}\rangle$=449+/-112 G in 2008 for $\varphi $=0.43. Zeeman
S-wave in V-spectrum smoothed over 3 {\AA} and its correspondence
to the dI($\lambda )$/d$\lambda $ wave is presented on figure 2b.

To find He II 4686 {\AA} line formation regions we constructed
Doppler tomogram (the binary system image in velocity space) on
the base of our VLT observational data. We used new Doppler
tomogram reconstruction technique worked out by Agafonov $^{24}$,
so-called Radioastronomical Approach (RA). This RA method includes
an effective CLEAN procedure and allows to reconstruct well the 2D
velocity field with very small number of 1D profiles (5-10 spectra
may be sufficient), see as example the figure 1 in $^{25}$. The
tomography map constructed on the base of all 13 HeII$\lambda
$4686 {\AA} line VLT profiles obtained by us in 2007 and 2008 is
presented in figure 3. It shows that HeII$\lambda $4686 {\AA} line
emission regions are located near the point L1 in the Roche lobe
model and near the "hot spot'' or "hot line" $^{26}$ on the outer
part of the accretion structure. But for the different
observational seasons Cyg X-1 tomogram maps may differ from one
another -- matter flow changes at the scale of years $^{27-28}$.
Therefore we constructed also tomograms using our VLT-profiles
separately for 2007 and 2008. They show a similar result.
Consequently $\langle B_{z}\rangle$ derived from HeII emission
line is located in outer parts of accretion structure. Its values
400-800 G are in agreement with estimation $^{15}$.

Our main conclusions are the following: We discovered a
longitudinal magnetic field averaged over the picture plane
$\langle B_{z}\rangle\sim 100$ G in the photosphere of Cyg$\sim
$X-1 optical component. Real magnetic field can exceed $\langle
B_{z}\rangle$. Magnetic field was detected at high confidence
level (about 6$\sigma )$ near orbital phase $\varphi $=0.4 in
2007, and $\varphi $=0.5 in 2008. Dependence of $\langle
B_{z}\rangle$ from $\varphi $ is more complicated than for
magnetic dipole and is probably changed during one year (figure
1). The magnetic field structure variation may be the reason of
some long-time variations of matter flowing process in this binary
system.

Further we found $\langle B_{z}\rangle\sim 600$ G in outer parts
of the accretion structure surrounding the BH. It is in agreement
with Shvartsman's ideas $^{3}$, that gas stream carries the
magnetic field to the accretion structure and the gas is
compressed by a factor of $\sim $10 due to the interaction with
the structure of the outer rim. Gas density is increased and
magnetic field is increased up to B $\sim $ 600 G at a distance
from BH $6\times 10^{11}$ cm = $2\times 10^{5}\times R_{g} \quad
^{29 }$ (R$_{g}$ is gravitation radius). According to
Shakura-Sunyaev $^{30}$ magnetized accretion disc standard model
at $3\times R_{g}$ B $\sim $ 10$^{9}$ G. Taking into account
radiative pressure predominance inside $\sim $10---20 R$_{g}$, we
get B($3\times R_{g} )\sim $(2---3)$\times 10^{8}$ G. The measured
value of the magnetic field strength at the marginal orbit of the
Cyg X-1 black hole corresponds quite well to the Magnetic Coupling
model with equipartition between kinetic and magnetic energy
densities.

If the X-ray millisecond flickering is related to the magnetic
nature, then the accreting matter magnetic energy flux must exceed
the X-ray emission fluctuating component luminosity. X-ray
emission originates at $R < 30\times R_{g}$. Inside the sphere of
this radius the magnetic energy amounts to 10$^{40}$ erg. The
radial velocity of magnetized plasma at $30\times R_{g}$ in
Shakura-Sunyaev accretion disk is $\sim $1.5 km/s (we adopt
viscosity parameter $\alpha $ = 1, because magnetic viscosity is
big $^{31-32})$. The time of matter fall is $\sim $1000 s.
Magnetic energy flux is 10$^{37}$ erg/s which is equal or exceed
the flickering component power $(0.5-1)\times 10^{37}$ erg/s. So,
magnetic energy dissipation permits to account for the X-ray
flickering.

From abovementioned estimations the Cyg X-1 black hole magnetic
moment is about 10$^{30} G\times cm^{3}$. According to $^{33}$
such object belongs to Magnetic Extremely Compact Object (MECO)
class.

So VLT FORS1 2007-2008 observations permit to detect the presence
of a magnetic field in Cyg X-1. It is the pioneer measurement in
black hole systems. The field can be responsible for X-ray
millisecond flickering. Our result points to necessity of taking
into account of magnetic field impact on the matter fluid
structure in Cyg X-1.

\section*{Acknowledgments}

This work has been partially supported by grants of Russian Foundation for
Basic Research and European South Observatory grants for VLT observations in
service mode. We thank Marcus Schoeller for discussion.

\bigskip

\section*{Figure captions}

\textbf{Figure 1. The mean longitudinal magnetic field of the Cyg
X-1 optical component }$\langle B_{z}\rangle$ \textbf{ (in Gauss)
vs. the orbital phase $\phi $. }Full squares: 2007 data; full
triangles: 2008 data. Error bars are 68{\%} confidence intervals.
The phases of orbital period 5.6 days were calculated with
ephemeredes $^{21}$, $\phi $ = 0 corresponds to the time, when BH
component is located behind the optical component (O9.7 Iab
supergiant). \textbf{a,} The graph for 2007. \textbf{b, }The graph
for 2008. \textbf{c, }The graph for 2007 and 2008 together.

\medskip

\noindent \textbf{Figure 2. Examples of Zeeman S-waves. }Two
examples showing accordance of wavelength dependence of intensity
normalized to continuum I/I$_{c}$ to Zeeman S-waves of V/I spectra
are given. \textbf{a,} An example for absorption spectral lines
originated in atmosphere of O-star component of Cyg X-1 binary
system: Zeeman S-wave (down box) in the region of HeI $\lambda
$4026{\AA} line (upper box) observed July 9, 2007 when $\langle
B_{z}\rangle$ was high (see table 1). \textbf{b, }Evidence of
magnetic field in accreting gas following from HeII $\lambda
$4686{\AA} emission line analyzes for June 18, 2007. The line on
the upper panel shows the observed HeII $\lambda $4686{\AA}
spectral line profile I/I$_{c}$. Horizontal line is
pseudo-continuum level. Solid line on the down panel shows V/I
spectrum smoothed over 3 {\AA}. The dashed line on the panel shows
in arbitrary units the expected Zeeman S-wave shape (dI/d$\lambda
)$/I (V/I $\sim $ (dI/d$\lambda )$/I), where I was smoothed over 3
{\AA}. Vertical solid line shows the centre of HeII $\lambda $4686
{\AA} emission.

\medskip

\noindent \textbf{Figure 3. Doppler tomogram of Cyg X-1 in HeII
$\lambda $4686 {\AA}.} It is constructed on the base of 13 VLT
spectra from 2007-2008 and shows brightness of emission in HeII
$\lambda $4686 {\AA} spectral line on velocities plane
(V$_{x}$,V$_{y})$. Thin lines are isophots of different levels
shown on the right part of figure (in relative units). The zero
level corresponds to ``pseudo-continuum''; negative values
correspond to absorption and positive ones -- to emission. The
dashed line shows the Roche lobe of black hole (BH). The almost
filling its Roche lobe optical component is drown by solid line.
The Roche lobes are constructed for the mass ratio q =
M$_{X}$/M$_{O}$ = 1/3. Here M$_{X}$ is the mass of X-ray emitting
component (BH) and M$_{O}$ is the mass of optical component
(O-star). Ovals represent outer parts of the accretion disks with
radii r$_{d}$ = 0.2 and 0.25 of the distance between mass centers
of the components.

\end{document}